# Directional Enhanced Probe for Side-Illumination Tip Enhanced Spectroscopy

*Hongming Shen,[1] Guowei Lu,[1,2,]\* Zhengmin Cao,[1] Yingbo He,[1] Yuqing Cheng,[1] Jiafang Li,[3] Zhi-Yuan Li,[3] and Qihuang Gong,[1,2]*

\*Corresponding Author: E-mail: guowei.lu@pku.edu.cn

[1] State Key Laboratory for Mesoscopic Physics, Department of Physics, Peking University, Beijing 100871, China
[2] Collaborative Innovation Center of Quantum Matter, Beijing 100871, China
[3] Laboratory of Optical Physics, Institute of Physics, Chinese Academy of Sciences, Beijing 100190, China

**Abstract** We demonstrate a high-performance apertureless near-field probe made of a tapered metal tip with a set of periodic shallow grooves near the apex. The spontaneous emission from a single emitter near the tip is investigated systematically for the side-illumination tip enhanced spectroscopy (TES). In contrast with the bare tapered metal tip in conventional side-illumination TES, the corrugated probe not only enhances strongly local excitation field but also concentrates the emission directivity, which leads to high collection efficiency and signal-to-noise ratio. In particular, we propose an asymmetric TES tip based on two coupling nanorods with different length at the apex to realize unidirectional enhanced emission rate from a single emitter. Interestingly, we find that the radiation pattern is sensitive to the emission wavelength and the emitter positions respective to the apex, which can result in an increase of signal-to-noise ratio by suppressing undesired signal. The proposed asymmetrical corrugated probe opens up a broad range of practical applications, e.g. increasing the detection efficiency of tip enhanced spectroscopy at the single-molecule level.

## 1. Introduction

Nanoscale chemical analysis for inhomogenous surfaces or interfaces is becoming more and more important due to advances and new challenges in areas such as heterogeneous catalysis, molecular electronics, and biology.[1] A microscopic technique which relies on the enhanced electric field near a sharp and laser irradiated metal tip, i.e. tip enhanced spectroscopy (TES), is the combination of scanning probe microscope (SPM) technique for high spatial resolution and optical spectroscopy for probing the chemical signatures of molecules. It is also known as a kind of "apertureless" version of scanning near-field optical microscopy.[2] The flexibility of the technique allows the study of various spectroscopic signals, including one- or two-photon excited fluorescence,[3] coherent anti-Stokes Raman scattering,[4] spontaneous Raman scattering, harmonic generations, and time-resolved measurements etc.[5-7] Specifically, since the experimental realization of tip enhanced Raman scattering (TERS), TERS imaging experiments have become possible in recent years on single molecules, graphene, carbon nanotubes, and biological samples.[8-12]

With TERS gaining increasing importance as a chemical analysis technique and being commercialized by an incremental number of manufacturers, questions about the rapid analysis of TERS imaging measurements arise. For instance, to obtain an image of 256×256 pixels with acquisition time at 0.5 s/pixel for weak Raman scatterers, it will take ~9 hours. For such a long time, the thermal drift of the SPM system can have a significant influence on the resulting image quality, and possible photodegradation of sample molecules can also limit the total imaging time. This calls to clarify whether the collection time per pixel can be reduced as much as possible to avoid the mapping time of several hours and the distortion of images. As a consequence of the inherently low Raman cross-section of most sample molecules, the excitation and detection of the scattered light has to be as efficient as possible to allow short acquisition time.[13] Thus, highly efficient detectors and optical systems, and strongly enhanced metallic apertureless tips are needed. Nowadays, optical detectors with single-photon detection capabilities are available, and current commercial optical systems present an optimized efficiency. Hence, these two factors are unlikely to provide enough space to reduce the collection time of the TERS imaging at orders of magnitudes.

The tip apex is no doubt the key device in TES as a local-scatter, which enhances and scatters the localized evanescent field towards the far-field. The near-field enhancement near nanoscale metal structures plays a central role in many nanoscale optical phenomena in plasmonics. The principle of TERS is related to surface-enhanced Raman scattering (SERS), where metallic nanoparticles or nanostructures lead to a large enhancement of the normally weak Raman signal by several orders of magnitude. For TERS, a metallic or metallized tip is illuminated by a focused laser beam and the resulting strongly enhanced electromagnetic field at the tip apex acts as a highly confined light source for Raman spectroscopic measurements. Up to now, a variety of optical probes have been proposed and applied to increase the spectroscopic response within a small sample volume.[14-17] However, the excitation enhancement effect for single tapered metal tip or gap-antenna between the apex and metal substrate is limited due to intrinsic nonradiative decay of the material and quantum tunneling effect. Despite much progress, most of the previous efforts are concentrated on the localized surface plasmon (LSP) resonance effect or lighting-rod effect.[18,19]

However, less attention has been devoted to the directivity of light emission modulated by the metallic tips,[20] which would provide another way to improve the TES signal greatly. Recently, optical antennas have been utilized to manipulate the emission direction of single molecules, which may provide an elegant way to realize much shorter acquisition time.[21-23] In our previous paper,[24] we reported on a three-dimensional corrugated metal probe with an unparalleled beaming effect of spontaneous emission from a single emitter, which improved the collection efficiency 25-fold compared to the conventional bare tips. In that work, the transmission-mode TES configuration was investigated and it is limited to transparent samples. To go beyond the limit, we present a theoretical extension to the side-illumination TES scheme. Although the side-illumination TERS with a metal substrate often provides highly efficient near-field enhancement, the low efficiency of the excitation and collection is an obvious weakness. Here, an asymmetric corrugated probe is proposed specifically for unidirectional and enhanced spontaneous emission, which is able to improve both the excitation and collection efficiency greatly. Numerical simulations show that the proposed tip offers larger near-field enhancement factors and more effective directional control in comparison with the conventional bare tapered probe. Such antenna design leads to an increase of the signal-to-noise ratio (SNR) greatly, which can reduce the collection time of TERS imaging over one order of magnitude. The proposed probe indicates a promising route to the development of directional enhanced emission related techniques.

## 2. Theoretical Methods and Models

Figure 1(a) illustrates the schematics of our proposed corrugated probe for TES. In side-illumination situation, a linear TM polarized beam is focused onto the tip/sample gap by a long working distance microscopic objective with an incidence angle of 60°. The strong local field at the apex can effectively excite nearby single emitters or scatters. The enhanced spontaneous emission from such perpendicular oriented emitter is collected using the same objective lens. In our calculations, the gold tip is modeled as a conical taper with an opening angle of 30° and tip-end radius $r = 10$ nm. To improve the collection efficiency, the geometrical parameters are required to be optimized carefully to enable more light directed towards the objective as shown in Fig. 1(b). Here we took the optimized groove period $G = 280$ nm, width $a = 140$ nm, depth $d = 40$ nm, groove number $N = 6$, and the tip-apex-to-first-groove distance $g = 280$ nm, respectively. In the experiment, the probe can be fabricated by current nano-fabrication methods, such as focused ion beam (FIB) milling and 3D direct laser writing methods.[25]

The finite-difference time-domain (FDTD) method is applied to simulate the emission process of a single dipole source near the metal tip.[26] The method has been widely used to calculate electromagnetic field distribution, scattering and absorption spectra, charge density distributions, and decay rates in the proximity of metallic nanostructures. We implement a classical point current source placed directly under the tip end in the FDTD calculations. The dielectric permittivity of gold is taken from the Johnson and Christy data and fitted by the Drude-Lorentz dispersion model.[27] Perfectly matched layers (PML) techniques are implemented as the absorbing boundary conditions. In the following, we set the mesh pitch to 2 nm unless stressed.

To obtain the far-field emission distribution, we first recorded the near-field data (**E**(t), **H**(t)) at two transformation surfaces $S_L$ and $S_R$, as depicted in Fig. 1(b), using FDTD simulations in the cylindrical coordinates. Because of the azimuthal symmetry, we can calculate the 3-dimensional directivity using 2-dimensional simulation results. Then we performed Fourier transform on these fields and calculated the surface electric (**J**$_s$) and magnetic (**M**$_s$) currents on the chosen surfaces $S_L$ and $S_R$, respectively. According to Near-Field to Far-Field (NTFF) transformation method,[28] we finally derived angular directivity $D(\theta) = \pi P(\theta) / \int P(\theta) d\theta$ for our 2-dimentional simulation results,[29] where $P(\theta)$ is the angle-dependent time-average Poynting power density and the integral is performed over the upper half space. To guarantee the accuracy of the simulation results, we have to point out that the transformation surfaces should be sufficiently large (here we choose $S_L$ and $S_R \sim 9$ μm long) and very much close to the dipole source or metal tip, which can capture most of the fields that contribute to the far-field emission.

## 3. Results and discussion

The side-illumination TES presents several advantages with respect to transmission-mode such as stronger excitation enhancement and not limited to transparent samples. However, since only a long working distance objective with low numerical aperture (less than 0.6 NA) can be used in the side-illumination configuration, two issues need to be considered carefully: (i) the focal spot projected on the sample surface is usually very large, as sketched in Fig. 1(a), leading to strong optical background noise; (ii) the collection efficiency of the system is limited by the small detection angle of the low NA objective lens. As a consequence, the side-illumination TES approach always suffers low SNR and collection efficiency.[10]

Our motivation is to improve the performance of the side-illumination TES, which is helpful for the single molecule detection with less collection time. We first consider the local field intensity in the excitation process based on 3D-FDTD simulations. In the calculation, the structured tip is under side-illumination by a *z*-polarized plane wave at the wavelength of 785 nm, as shown in Fig. 2(a) (It should be noted that the optimized response wavelength is tunable by the parameters of the corrugated probe). The Yee cell size is set to be $5 \times 5 \times 5$ nm$^3$. Figures 2(b) and (c) demonstrate the dominant *z*-component electric field distribution $|E_z|^2/|E_{in}|^2$ in the *x-z* plane for probe without and with the gratings, respectively. $E_{in}$ is the incoming excitation field amplitude. For both cases, we observe a strong and highly localized field at the tip apex due to the lighting-rod effect of a sharp tip. In the case of the corrugated tip, the maximum field enhancement *M* reaches up to 3755-fold, which is 25 times higher than the bare case (148-fold). The corrugated gratings provide necessary momentum matching condition for efficient surface plasmon polaritons (SPPs) coupling.[23] Figures 2(d) and (e) plot the normalized curves of local field intensity along line *z* and *x* direction, respectively. The curve profiles of both cases are almost the same, indicating the gratings can strongly enhance the field intensity but do not affect its spatial distribution (~ 18 nm) which is related with the spatial resolution of the TES probe.[24]

Figure 3 demonstrates the calculated angular directivity of the emission obtained with the NTFF method. A *z*-oriented dipole emitter at 670 nm (e.g. corresponding to emission band of widely used Cyanine 5 dye molecule or fluorescent nanodiamond) is placed at a distance of 10 nm above the Au substrate. We compare three cases: (i) an isolated dipole emitter above the Au substrate, (ii) an emitter coupled to a tip without and (iii) with corrugated gratings. We assumed a 0.5 NA objective with maximum detectable angle of ±30° fixed at angle of 60° with respect to the tip axis. To quantify the collection efficiency of the system, we define the factor $\kappa = P_1/P_2$, where $P_1 = \int_0^{\frac{\pi}{3}} P(\theta)d\theta$ is the power within the detectable angle of the objective and $P_2 = \int_0^{\pi} P(\theta)d\theta$ is the total power radiated to the upper half space for the 2-dimensional simulation results in the *x-z* plane. It should be noted that the factor $\kappa$ defined here is a simple approximation of the collection efficiency for comparison, in practice, the integration should be considered in the full solid angle.[29] For an isolated dipole emitter close to the Au substrate, it exhibits a typical "doughnut" radiation pattern. Such emission spreads over the collection angles with the factor $\kappa = 0.45$, resulting in large background noise. Now we consider a bare tip with total length of $L = 2000$ nm located 10 nm away from the dipole emitter. It can be seen that the angle of the main beam $\theta_m$ is larger than 60° and only a small part of the emitted power can be collected by the objective ($\kappa = 0.18$).[30,31] This often leads to a low SNR for the conventional bare TES tips. By introducing the concentric metal gratings, we observe an obvious emission beaming effect and the angle $\theta_m$ is tuned towards 30°, thus the emitted power can be easily captured by the objective. The factor $\kappa$ of the corrugated tip is 0.47, which is 2.6-fold higher than the bare one. Such directional emission is due to a complex interference of the light emitted directly by the dipole source and surface waves scattered by the surface corrugated grooves towards the far-field, which has been demonstrated in the previous work using the Green's function formalism.[24]

Figure 4 plots the dependence of the factor $\kappa$ on the emission wavelength λ. It can be seen that the bare tip has low collection efficiency ($\kappa \sim 0.2$) with varying λ values from 600 to 800 nm. Compared to the bare tip, the emission pattern with the corrugated tip is strongly dependent on the wavelengths, as shown in Figure 4(a)-(d). From λ = 500 to 800 nm, dramatically enhanced collection efficiency ($\kappa > 0.4$) can be obtained for a wide range of wavelengths due to the beaming effect of the surface corrugated gratings. Remarkably, for the emission wavelength at 532 nm, as shown in Fig. 4(b), the light is highly directed into a single narrow beam with $\theta_m$ at 45° and a full-width at half-maximum (FWHM) of 30°. While further increasing λ, for example, λ = 900 nm, the factor $\kappa$ decrease rapidly to 0.24 and the emission far-field pattern is quite similar to the bare tip. One also notes that $\kappa$ reaches up to 0.49 for the bare tip with emission pattern similar to the isolated dipole case at λ = 500 nm, which is ascribed to the weak coupling of antenna-dipole system. The broadband of enhanced directivity makes the antenna design more convenient for nanoscale optical spectroscopy applications, for example, fluorescent molecules with large anti-Stock shifts between laser and emission wavelength, and two-photon fluorescence.

Efficient detection of fluorescence or Raman signal is highly desirable and challenging in single-molecule detection and analytics.[8] However, the emission patterns discussed above is omnidirectional in the azimuth due to the symmetry of the antenna. This would lead to more than half of the signal lost. Inspired by Pakizeh and Kall,[22] unidirectional emission can be realized if an emitter coupled to a dark mode induced in two neighboring metallic disks. Here, taking into account of the scanning probe microscope technique, we proposed to place two asymmetrical nanorods at the tip end, i.e. one rod much longer than the other, avoiding to scan surface with two apexes. And the shorter rod would couple with the surface emitters weakly with compared to the longer one. The conceptual device structure is sketched in the Figure 5(a). The asymmetric probe can be decomposed into three main parts: a sharp metal probe with tip radius of 25 nm and opening angle of 30°, several grooves milled on one side of the tapered metal tip ($G = g = 390$ nm, $a = 185$ nm, $d = 30$ nm, and $N = 5$), and a dimer resonator (rod radius $r = 10$ nm, length $D_1 = 80$ nm, $D_2 = 40$ nm, and surface-surface distance $\delta = 8$ nm) carved at the tip end. Fig. 5(b) shows the near-field distribution for the asymmetric corrugated tip. It can be seen that a strong electromagnetic field is highly confined in the gap between the tip end and metal surface. The dependence of the angular directivity on the emitter's position is explored in Figure 5(c)-(d). As for point c (Fig. 5(c)), a *z*-oriented dipole at wavelength of 800 nm is placed 4 nm directly under the longer nanorod. The radiation patterns demonstrate excellent unidirectional emission for both asymmetric tips without and with the surface corrugated gratings. Such unidirectional effect is mainly due to the near-field coupling interaction of phase shifts between the two rods' dipole mode.[22] While the gratings here are responsible for further beaming light, resulting in a higher collection efficiency with the factor $\kappa = 0.83$ for the corrugated asymmetrical tip. Compared to the conventional TES tip ($\kappa = 0.18$), the factor $\kappa$ of the asymmetric corrugated tip is almost 5-fold higher. Varying point c to g, the unidirectionality effect diminishes and more light is directed towards the left side. Interestingly, one may note that the direction of the main radiation lobe is reversed to the left side if the emitter is positioned near the shorter nanorod at point g (Fig. 5(g)). This would suppress the background signal, leading to further increasing of the SNR of the fluorescence or Raman signal, which is helpful for sensitive single-molecule detection.

## 4. Conclusion

In summary, we have investigated the emission properties of a 3D corrugated metal probe with several concentric gratings for the side illumination TES scheme. The simulation results show that the angular emission from a single emitter can be highly directed into a single narrow lobe at particular polar angle, which is tunable by the antenna parameters and also wavelength dependent. Such probe simultaneously increases the localized field intensity to be about 25 times stronger than conventional bare tip. Remarkably, an asymmetric TES tip with the asymmetrical dimer resonator is proposed to further improve the collection efficiency. The collection efficiency is almost 5-fold higher than the conventional TES tip due to the unidirectional beaming effect. We also explored the dependence of the angular directivity on the emitter's position, interestingly, the emitters away from the tip apex present different emission direction compared to ones below the apex, which implies lower unspecific background. The highly tunable emission properties of the proposed tip would lead to an increase of the signal-to-noise ratio and less collection time, providing a promising approach for fundamental applications in optical spectroscopy, analytical chemistry, single-molecule sensing, and tip enhanced fluorescence and Raman imaging techniques.

**Acknowledgements**

This work was supported by the National Key Basic Research Program of China (grant nos.2013CB328703) and the National Natural Science Foundation of China (grant nos. 61422502, 11374026, 91221304, and 11431017).

**Figure 1**

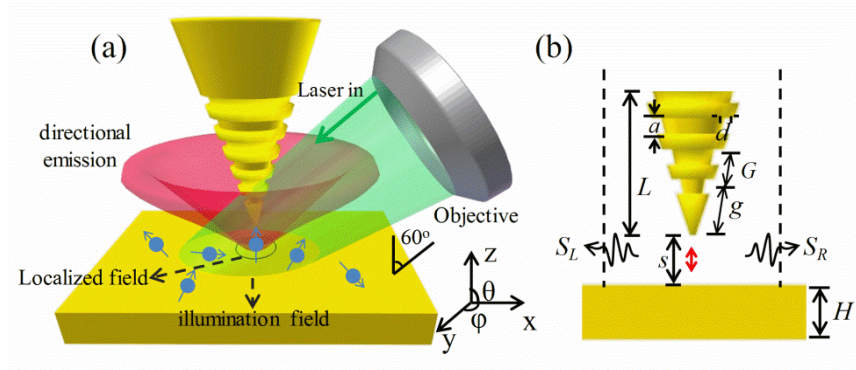

**Figure 1.** (a) Schematics of the proposed side-illumination/collection TERS configuration. A linear TM polarized beam passes through a long working distance objective with an incident angle of 60°. (b) Sketch of the metallic tip design: $G = g = 280$ nm, $a = 140$ nm, $d = 40$ nm, $N = 6$. The radius of curvature at the tip end is 10 nm with an opening angle of 30°. The separation $s$ between the tip and the gold substrate (thickness $H = 100$ nm) is 20 nm, and a $z$-oriented dipole emitter is placed in the middle.

**Figure 2.**

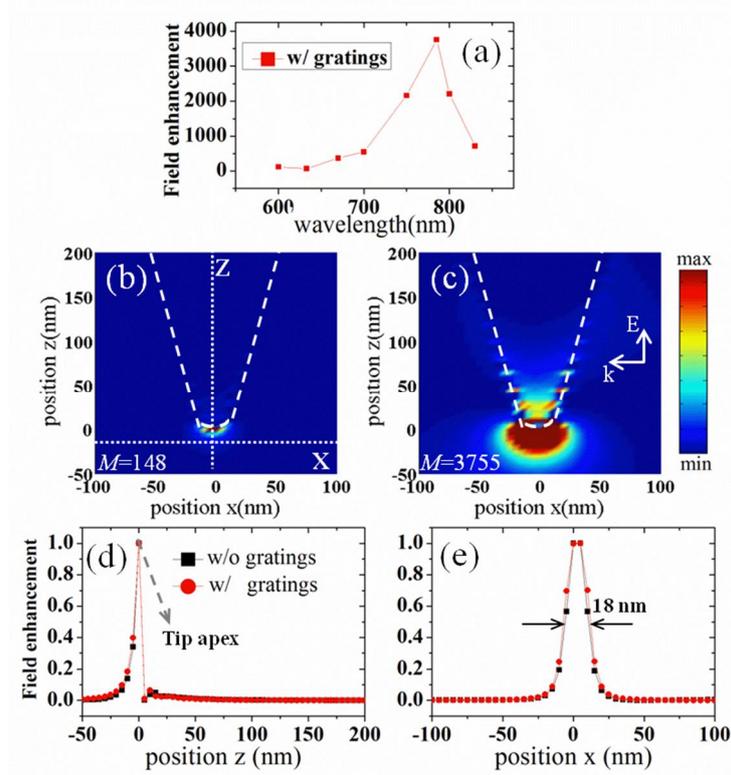

**Figure 2.** Localized excitation enhancements based on 3D-FDTD simulations. (a)Field enhancement $|E_z|^2/|E_{in}|^2$ as a function of wavelength. (b) and (c) demonstrate the electric field distribution for tip without and with gratings in free space, respectively, while (d) and (e) plot the normalized curves of local field enhancements along line profile $z$ (along the tip axis) and $x$ ( at a surface 5 nm under the tip apex) direction. For the corrugated tip, the maximum field enhancement $M$ at the tip apex is 3755-fold, which is 25 times higher than the bare one.

**Figure 3.**

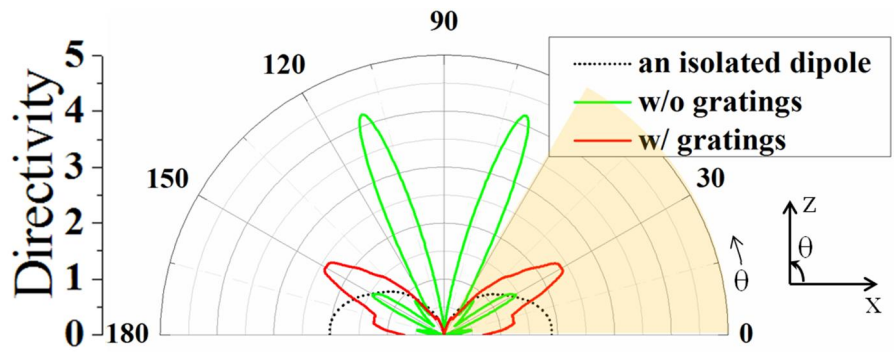

**Figure 3.** Calculated far-field radiation patterns in the *x-z* plane of a *z*-oriented dipole ($\lambda$ = 670 nm) close to the metal substrate for three cases: an isolated dipole emitter (black dot line), an emitter coupled to a tip without (green solid line) and with gratings (red solid line), respectively. The coordinate system and definition $\theta$ are also shown in the figure.

**Figure 4.**

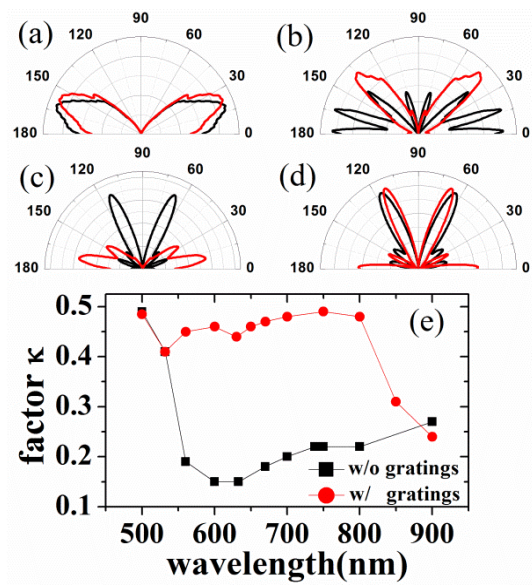

**Figure 4.** (a)-(d) Calculated radiation patterns for emission wavelength at 500 nm, 532 nm, 800 nm, and 900 nm for tip without (black) and with (red) gratings, respectively. (e) Dependence of the factor $\kappa$ on the dipole emission wavelength. Taking advantage of plasmonic beaming, the corrugated tip exhibits much higher collection efficiency ($\kappa > 0.4$) for a wide range of wavelengths as compared to the bare one ($\kappa \sim 0.2$).

**Figure 5.**

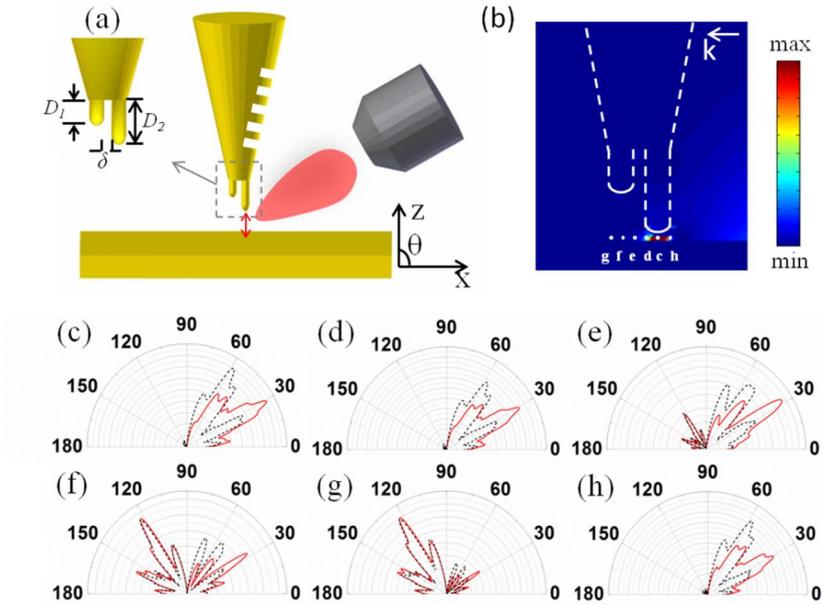

**Figure 5.** Hybrid asymmetric probe for unidirectional single-molecule emission by 2D-FDTD simulations in the Cartesian coordinates. (a) Schematics of the hybrid asymmetric probe. (b) Calculated near-field distribution $|E_z|^2/|E_{in}|^2$ in the *x-z* plane of the asymmetric probe coupled to the Au substrate. In the calculations, a *z*-polarized plane wave at 565 nm is incident from +*x* direction and the separation between the tip and Au substrate is 4 nm. (c)-(h) show the position dependence of the angular patterns for single emitters placed at c(0, -4), d(-10, -4), e(-20, -4), f(-28, -4), g(-40, -4), h(10, -4) in nm, respectively, while the tip apex is fixed at (0, 0) nm. The red line plots for the asymmetric tip with gratings and black dashed line for the bare tip.

**Graphical Abstract**

Directional and enhanced spontaneous emission controlled by an asymmetric corrugated metal probe is demonstrated to improve the detection sensitivity of tip enhanced spectroscopy.

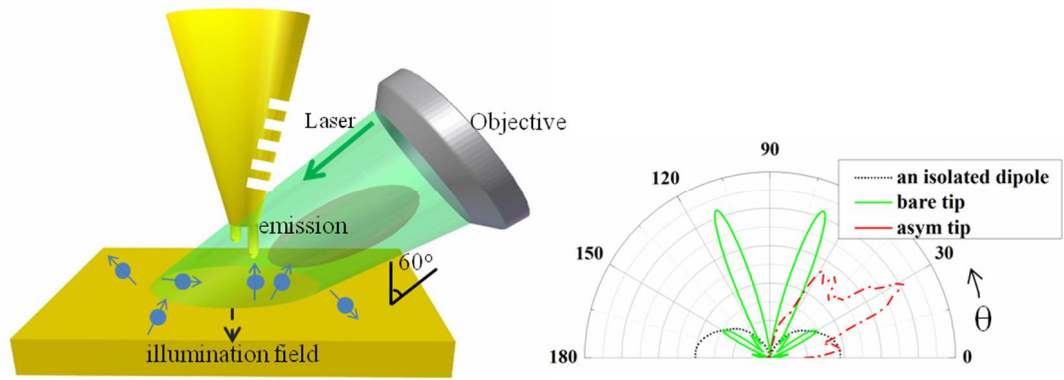

# References

[1] E. A. Pozzi, M. D. Sonntag, N. Jiang, J. M. Klingsporn, M. C. Hersam, R. P. Van Duyne, *ACS Nano* **2013**, *7*, 885-888.
[2] E. Bailo, V. Deckert, *Chem. Soc. Rev.* **2008**, *37*, 921-930.
[3] A. Hartschuh, M. R. Beversluis, A. Bouhelier, L. Novotny, *Phil. Trans. R. Soc. Lond. A* **2004**, *362*, 807-819.
[4] T. Ichimura, N. Hayazawa, M. Hashimoto, Y. Inouye, S. Kawata, *Phys. Rev. Lett.* **2004**, *92*, 220801.
[5] T. Schmid, L. Opilik, C. Blum, R. Zenobi, *Angew. Chem. Int. Ed.* **2013**, *52*, 5940-5954.
[6] B. Pettinger, P. Schambach, C. J. Villagómez, N. Scott, *Annu. Rev. Phys. Chem.* **2012**, *63*, 379-399.
[7] A. D. Elfick, A. Downes, R. Mouras, *Anal. Bioanal. Chem.* **2010**, *396*, 45-52.
[8] R. Zhang, Y. Zhang, Z. C. Dong, S. Jiang, C. Zhang, L. G. Chen, L. Zhang, Y. Liao, J. Aizpurua, Y. Luo, J. L. Yang, J. G. Hou, *Nature* **2013**, *498*, 82-86.
[9] Z. D. Schultz, J. M. Marr, H. Wang, *Nanophotonics* **2014**, *3*, 91.
[10] M. Zhang, R. Wang, X. Wu, J. Wang, *Sci. China-Phys. Mech. Astron.* **2012**, *55*, 1335-1344.
[11] W. Zhang, B. S. Yeo, T. Schmid, R. Zenobi, *J. Phys. Chem. C* **2007**, *111*, 1733-1738.
[12] X. Wang, Z. Liu, M. Zhuang, H. Zhang, X. Wang, Z. Xie, D. Wu, B. Ren, Z. Tian, *Appl. Phys. Lett.* **2007**, *91*, 101105.
[13] J. Stadler, T. Schmid, R. Zenobi, *Nano Lett.* **2010**, *10*, 4514-4520.
[14] N. Mauser, A. Hartschuh, *Chem. Soc. Rev.* **2014**, *43*, 1248-1262.
[15] A. Bek, F. De Angelis, G. Das, E. Di Fabrizio, M. Lazzarino, *Micron* **2011**, *42*, 313-317.
[16] C. Höppener, Z. J. Lapin, P. Bharadwaj, L. Novotny, *Phys. Rev. Lett.* **2012**, *109*, 17402.
[17] M. Mivelle, T. S. van Zanten, M. F. Garcia-Parajo, *Nano. Lett.* **2014**, *14*, 4895-4900.
[18] C. Ropers, C. C. Neacsu, T. Elsaesser, M. Albrecht, M. B. Raschke, C. Lienau, *Nano Lett.* **2007**, *7*, 2784-2788.
[19] Z. Yang, J. Aizpurua, H. Xu, *J. Raman Spectrosc.* **2009**, *40*, 1343-1348.
[20] A. Ghimire, E. Shafran, J. M. Gerton, *Sci. Rep.* **2014**, *4*, 6456.
[21] T. H. Taminiau, F. D. Stefani, F. B. Segerink, N. F. Van Hulst, *Nat. Photon.* **2008**, *2*, 234-237.
[22] T. Pakizeh, M. Käll, *Nano Lett.* **2009**, *9*, 2343-2349.
[23] H. Aouani, O. Mahboub, N. Bonod, E. Devaux, E. Popov, H. Rigneault, T. W. Ebbesen, J. Wenger, *Nano Lett.* **2011**, *11*, 637-644.
[24] H. Shen, G. Lu, Y. He, Y. Cheng, H. Liu, Q. Gong, *Nanoscale* **2014**, *6*, 7512-7518.
[25] J. Li, J. Mu, B. Wang, W. Ding, J. Liu, H. Guo, W. Li, C. Gu, Z. Li, *Laser Photonics Rev.* **2014**, *8*, No. 4, 602–609.
[26] A. F. Oskooi, D. Roundy, M. Ibanescu, P. Bermel, J. D. Joannopoulos, S. G. Johnson, *Phys. Commun.* **2010**, *181*, 687-702.
[27] P. B. Johnson, R. W. Christy, *Phys. Rev. B* **1972**, *6*, 4370-4379.
[28] A. Taflove, S.C. Hagness, Computational Electrodynamics: The Finite-Difference Time-Domain Method, Third Edition. Norwood, MA: Artech House, **2005**.
[29] Y. C. Jun, K. C. Y. Huang, M. L. Brongersma, *Nat. Commun.* **2011**, *2*, 283.
[30] T. H. Taminiau, F. D. Stefani, N. F. van Hulst, *Nano Lett.* **2011**, *11*, 1020-1024.
[31] R. Paniagua-Dominguez, G. Grzela, J. G. Rivas, J. A. Sanchez-Gil, *Nanoscale* **2013**, *5*, 10582-10590.